\documentclass[twocolumn,amssymb,showpacs,aps,prl]{revtex4-1}
\usepackage{}
\usepackage{stmaryrd}
\usepackage{amsmath}
\usepackage{amssymb}
\usepackage{float}
\usepackage{graphicx}
\usepackage{titletoc}
\usepackage{booktabs}
\usepackage{array, color}
\usepackage{tabularx}
\usepackage[colorlinks,linkcolor=blue,anchorcolor=blue,citecolor=blue,urlcolor=blue,dvipdfm]{hyperref}
\usepackage{subfigure}
\usepackage{latexsym,bm}
\usepackage{graphics}
\begin{document}

\title{Dynamic core polarization in strong-field ionization of CO molecules}

\author{Bin Zhang}
\author{Jianmin Yuan}
\author{Zengxiu Zhao}
\email{zhao.zengxiu@gmail.com}
\affiliation{Department of Physics, College of Science, National University of Defense Technology, Changsha 410073, Hunan, People's Republic of China}

\date{\today}

\begin{abstract}
The orientation-dependent strong-field ionization of CO molecules is investigated using the fully propagated three-dimensional time-dependent Hartree-Fock theory.
The full ionization results are in good agreement with recent experiments.
The comparisons between the full method and single active orbital (SAO) method show that although the core electrons are generally more tightly bounded and contribute little to the total ionization yields, their dynamics cannot be ignored, which effectively modify the behaviors of electrons in the highest occupied molecular orbital. By incorporating it into the SAO method, we identify that the dynamic core polarization plays an important role in the strong-field tunneling ionization of CO molecules, which is helpful for future development of tunneling ionization theory of molecules beyond single active electron approximation.
\end{abstract}


\maketitle


Tunneling ionization (TI) is one of the most prominent  processes for atoms and molecules subjected to strong laser fields. It ignites various atomic dynamics such as high-order harmonic generation (HHG), which forms the basis for attosecond science~\cite{krausz09,mcfarlan08,smirnova09}. Recent advances in  shaping and tailoring laser pulses as well as aligning molecules provides a better controlled manner to explore fundamental concepts involved in TI, e.g., tunneling time and tunneling wave packet \cite{Uiberacker07,Eckle08n,Eckle08s,Shafir12N}.
In general, the molecular ionization dynamics in strong fields can be successfully described by the quasi-static theories \cite{Tong02,Kjeldsen05,Kjeldsen05,muth00} which assume the molecular core is frozen and the laser field is not varying during the TI process.   Studies based on these single active electron (SAE) models concluded that the orientation-dependent ionization rate maps the asymptotic electron density distribution, leading to the imaging of the ionizing orbitals~\cite{Tong02,Kjeldsen05,muth00,alnaser05}.

However, it has been questioned recently when the observed ionization yields deviate from the expectation based on the shape of the highest occupied molecular orbital (HOMO)~\cite{pavicic07,murray11,akagi09,wu11,petretti10,Hol10,dimitrovski10,dimitrovski11}.
%
Among the various proposed models, the linear Stark effect \cite{Hol10,dimitrovski10,dimitrovski11} has been incorporated  into the molecular Ammosov-Delone-Krainov (MO-ADK) theory~\cite{Tong02,Kjeldsen05} to explain the orientation-dependent ionization of  the OCS molecule~\cite{Hol10}.
But recent measurements~\cite{wu12} on the orientation-dependent ionization of CO molecule deviate apparently from the Stark corrected MO-ADK.
 Other experiments~\cite{li11,ohmura11} also indicate that the linear Stark effect plays a minor role and the ionization rate is dominated by the orbital profile.
The numerical study based on SAE potentials~\cite{abu10} does not solve this puzzle on CO, which triggers the even challenging need for a non-perturbative treatment of the multi-electron dynamics of molecules in intense laser pulses.

It has been evidenced that multi-orbital and multi-dipole effects come into play for strong field physics
\cite{mcfarlan08, smirnova09, akagi09, Worner10L, Pabst12,patchkovskii06,zhao07pra}. However, the direct numerical integration of the time-dependent Schr\"odinger equation (TDSE) is computationally prohibitive for systems with more than two electrons~\cite{birkeland10,zhang11,guan11}.
The popular approximations include the time-dependent density-functional-theory (TDDFT)~\cite{telnov09,chu11,fowe11}, and
the multi-configuration time-dependent Hartree-Fock (MCTDHF) theory~\cite{caillat05,haxton11,hochstuhl11}, but they both suffer
from some disadvantages.

In this letter, we investigate the orientation-dependent strong-field ionization of CO in intense laser fields by the fully propagated three-dimensional time-dependent Hartree-Fock (TDHF) theory~\cite{kulander87}, within the Born-Oppenheimer approximation.
TDHF goes beyond the SAE approach and includes the response to the field of all electrons~\cite{kulander87}, which helps to identify the multipole effects from the molecular core in strong field TI.
The full ionization results are compared with the experiment~\cite{wu12}, and good agreements are reached. Furthermore, we have performed the calculations using the single active orbital (SAO) approximation, i.e., propagate the HOMO electrons while freezing the others.
The comparisons between the SAO method and full method show that although the core electrons are generally more tightly bounded and contribute little to the total ionization yields, their dynamics cannot be ignored, which effectively modify the behaviors of HOMO electrons. We demonstrate that the dynamic core polarization plays an important role in the strong-field tunneling ionization of CO molecule.

TDHF is a single determinant theory and may therefore be applied to quite large systems.
Although it includes no correlation, in strong-field cases, the question of how much and under what conditions correlation beyond the Hartree-Fock model is important still remains discussed~\cite{niko07,kaiser11,lotstedt12}.
For the numerical implementation, we use the prolate spheroidal (PS) coordinates~\cite{zhang12}, which is almost natural choice for two-center systems.
Our approach is also based on the discrete-variable representation (DVR) and the finite-elements method (FEM)~\cite{guan11}.
DVR offers distinct advantages in the representation of local potential operators, while FEM provides more flexibility in the design of numerical grid and increases the sparseness of the kinetics matrix.
For the temporal propagation, we use the efficient Short Iterative Lanczos (SIL) algorithm~\cite{park86}.

The numerical parameters are as follows.
The internuclear distance of CO is fixed at experimental equilibrium of 2.132 a.u.~\cite{nist}.
As the ground electronic state is $^1\Sigma$, spin-restricted form of TDHF is adopted here.
The ground state is determined by relaxing the system in imaginary time from a guess wavefunction.
The total (HOMO) energy of CO from relaxation calculation is -112.7909118 (-0.554923304) a.u., in good agreement with literature values~\cite{kobus93}.
The electric field ${\bf E}(t)$ is linearly polarized (in the xz-plane, with $\beta$ denotes the orientation angle with respect to the molecular axis), $E(t)=E_0\sin^2(\pi t/T)\sin(\omega t+\phi)$, where $E_0$ is the peak field amplitude, $\omega$ is the carrier frequency, $T$ is the pulse duration and $\phi$ is the carrier envelope phase (CEP).
The laser intensities of interest are in unit of $I_0=10^{14}$W/cm$^2$.
After the time propagation, we yield the total (orbital) wave function $\Psi(T)$ [$\psi_i(T)$].
The ionization probability from orbital $i$ is calculated as $p_i=1-\langle \psi_i(T)|\psi_i(T)\rangle$.
The total ionization probability $P=1-\langle \Psi(T)|\Psi(T)\rangle=1-\prod_i (1-p_i)$, which can be approximated as
$P\approx\sum_i p_i$ for small ionizations ($p_i\ll 1$).


\begin{figure}[t]
\centering
\includegraphics*[width=3in]{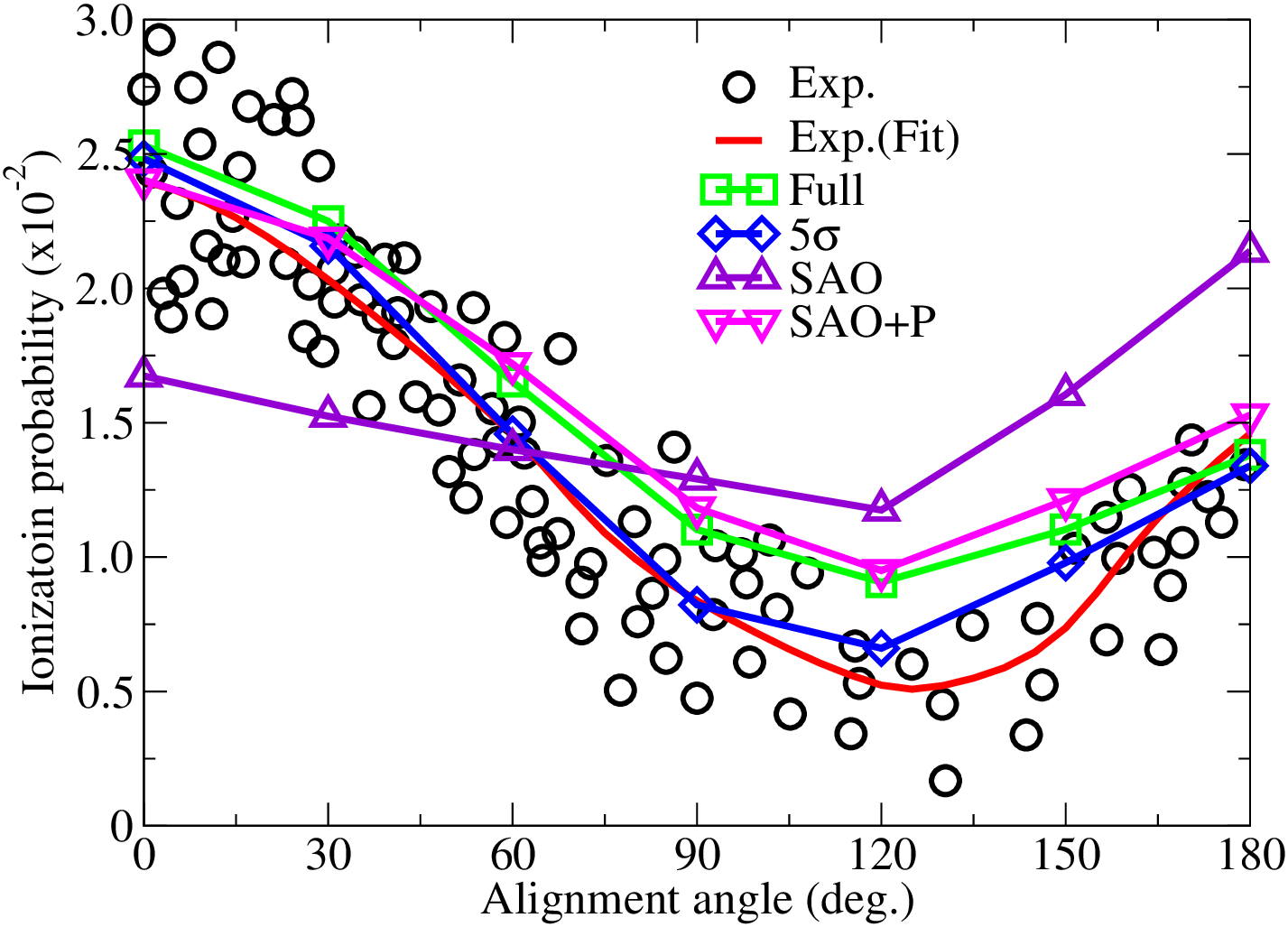}
\caption{(Color online)
The calculated ionization yields of CO versus the orientation angle $\beta$: total ($\square$) and 5$\sigma$ ($\Diamond$) (from full method), SAO method ($\triangle$) and SAO+P method ($\nabla$).
The experiment data (scattered $\circ$) are taken from ~\cite{wu12} (matched to the full calculation at $\beta=0^{\circ}$), where a circularly polarized laser field is used with estimated intensity of $4\times 10^{14}$W/cm$^2$ and pulse duration of 35-fs.  The fitted curve of experimental data is shown  in the red line.}
\label{iprob3t}
\end{figure}

The orientation-dependent ionization yields of CO have been measured in experiment~\cite{wu12},
where it's found that the CO molecule is easier to be ionized at parallel orientation than at anti-parallel orientation.
In experiment, a circularly polarized laser field with estimated intensity of 4$I_0$ and pulse duration of 35-fs was used.
For our linearly polarized laser fields, we use laser pulses of three optical cycles, and an equivalent intensity of 2$I_0$. The wavelength is 800nm and CEP $\phi=\pi/2$.
The O (C) atom is located at the negative (positive) part of the z-axis.
Thus, for the parallel orientation ($\beta=0^\circ$), the maximum laser electric field points from C to O, while it points from O to C for the anti-parallel orientation ($\beta=180^\circ$).
According to the tunneling theory~\cite{Tong02}, ionization rate decreases exponentially with the laser intensity.
As a result, the ionization induced by the field with peak intensity dominates the total ionization yields, filtering out the cycle-averaging effects.
Also, the ionization potential (IP) of 1$\pi$ (HOMO-1) is 2.3eV higher than that of 5$\sigma$~\cite{kobus93}, thus a small ionization from inner orbitals is expected.
The ionization yields are compared with experiment in Fig.~\ref{iprob3t}.
The full method yields a ratio $P(180^\circ)/P(0^\circ)=0.62$, in good agreement with the averaged ratio extracted from experiment.
Both the calculation and experiment predict an minimum around 120$^\circ$.
Our results show that the total ionization is dominated by $5\sigma$.
Due to the orbital symmetry, ionization from $1\pi$ is suppressed for both $\beta=0^\circ$ and $180^\circ$.
Observable contribution is available only around perpendicular orientations, where ionization from $5\sigma$ is suppressed.
Thus we conclude that the combination contribution of multiple orbitals plays a minor role.

In order to identify the core effects, we perform SAO calculations in which only $5\sigma$ electrons are active, in contrast to  the full method where all the electrons are fully propagated.
Note that in our SAO method, the direct and exchange potentials between the core and HOMO electrons are calculated exactly each time step in the framework of TDHF, which is different from SAE methods where local model potentials are used~\cite{petretti10,abu10}.
The orientation-dependent ionization yields from SAO calculations deviate obvious from the full calculations and experiment (Fig.~\ref{iprob3t}).
The SAO method predicts a larger ionization for $\beta=180^\circ$, yielding a ratio $P(180^\circ)/P(0^\circ)=1.28$, in qualitative
disagreement with experiment.
A previous TDSE study~\cite{abu10} based on a SAE potential  fails as well showing that a treatment beyond single active electron/orbital is required.

Hinted by the agreement of the experiment  with the full method which includes the responses of core electrons, we attempt to improve the SAO calculation by including the dynamic core polarization induced by the intense laser field~\cite{zhao07},
\begin{equation}
\label{vpol}
V_p({\bf r},t)=-\frac{\boldsymbol{\alpha}_c{\bf E}(t)\cdot {\bf r}}{r^3}
\end{equation}
where $\boldsymbol{\alpha}_c$ is the total polarizability of core electrons. The SAO method including $V_p$ is noted as SAO+P.
We calculate this polarizability in the following way.
In the full propagation, we have checked that the induced dipole moments of core electrons were mainly contributed from 1$\pi$ and 4$\sigma$ (HOMO-2). Fitting $\boldsymbol{\alpha}{\bf E}(t)$ to the numerical induced dipole moment ${\bf d}_{ind}(t)$ of each orbitals yield:
$\boldsymbol{\alpha}_{1\pi}$=(2.55,2.55,4.68) a.u. and $\boldsymbol{\alpha}_{4\sigma}$=(0.73,0.73,0.64) a.u.
The total polarizability can be approximated as $\boldsymbol{\alpha}_c\approx \boldsymbol{\alpha}_{1\pi}+\boldsymbol{\alpha}_{4\sigma}$.
Close to the core, we apply a cutoff for $V_p$, at a point where the polarization field
cancels the laser field~\cite{zhao07}. This is also necessary to remove the singularity near the core.
Taking the z-axis for example, the cutoff point $z_c$ satisfies $\alpha_{zz}E/z_c^2-z_cE=0$, which results in $z_c=\alpha_{zz}^{1/3}$.
As $\boldsymbol{\alpha}_c$ is anisotropic in general, all the cutoff points constitute an ellipsoidal surface.
The SAO+P results are in good agreement with the full calculations and experiment (see Fig.~\ref{iprob3t}).

By including the dynamic polarization, we see that $V_p$ enhances ionization for $\beta=0^\circ$, while ionization is suppressed for $\beta=180^\circ$.  This contradicts to the prediction of the static theory  which collaboates the polarization effects into an effective IP~\cite{Hol10}
\begin{equation}
\label{ipeff}
I_p^{\text{eff}}({\bf E})=I_p(0)+\Delta{\bf \mu}\cdot {\bf E}+\frac{1}{2}{\bf E}^T\Delta{\bf \alpha} {\bf E}
\end{equation}
where $\Delta{\bf \mu}$ ($\Delta{\bf \alpha}$) is the difference of the permanent dipole moment (polarizability) between CO and CO$^+$, $I_p(0)$ is the IP of CO in the absence of external fields.
Linear Stark shift takes the second term on rhs. of Eq.~(\ref{ipeff}) into consideration.
$I_p$ is raised (reduced) when the laser field is directed parallel (antiparallel) to the orbital dipole.
As a result, the linear Stark effect reverses the orientation-dependent ionization rate and indicates a maximum ionization for $\beta=180^\circ$~\cite{li11}.  It can be seen that the second order Stark shift correction [the third term on rhs. of Eq.~(\ref{ipeff})] is helpless in this situation since: $I_p$ is raised for both the $\beta=0^\circ$ and $180^\circ$. Therefore the direct inclusion of the polarizability term (static polarization) in IP [Eq.~(\ref{ipeff})] does not improve the MO-ADK theory in the CO case.

\begin{figure}[t]
\centering
\includegraphics*[width=3in]{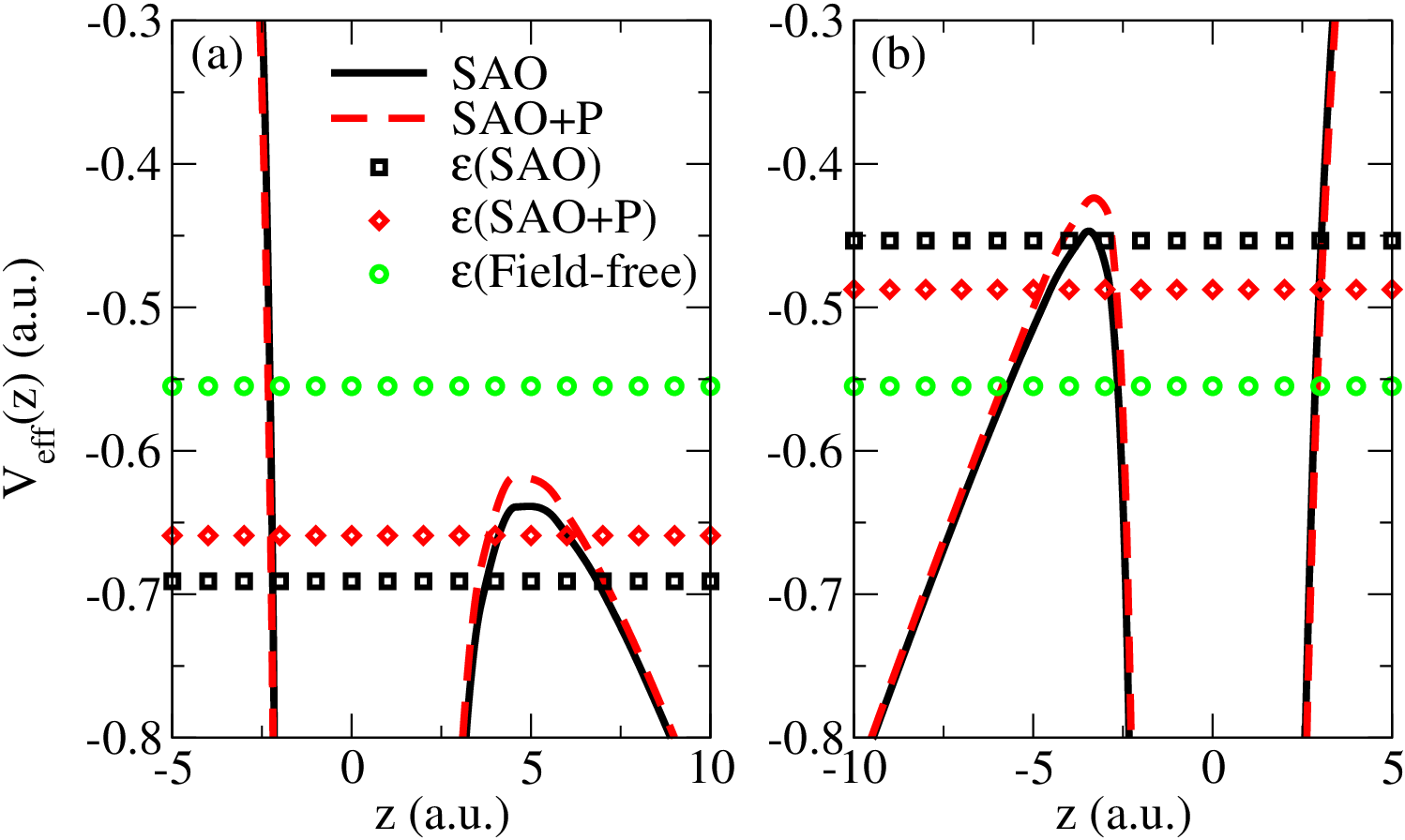}
\caption{(Color online)
Effective potential on the $5\sigma$ electrons along the molecular axis, predicted by the SAO method (solid lines) and SAO+P method (dashed lines), when the maximum laser electric field ($|E|=0.0755$ a.u.) points to O (a) and to C (b), respectively.
The horizontal marks represent the corresponding field-dressed and field-free orbital energies of $5\sigma$.
}
\label{veff}
\end{figure}

A theory of tunneling ionization in complex systems (CS-ADK)~\cite{zhao07,brabec05,zhang10cpl} has been proposed a few years ago, where a similar core polarization potential has been explicitly taken into account to improve the original MO-ADK theory.
We have calculated the ionization probabilities of CO by CS-ADK, but the results are similar to those from MO-ADK, except for a mild suppression of the probabilities for both $\beta=0^\circ$ and $180^\circ$.
To investigate the effects of $V_p({\bf r},t)$, we plot the effective potential $V_{eff}$ felt by the $5\sigma$ electron along the molecular axis (Fig.~\ref{veff}). $V_{eff}$ is defined as:
\begin{equation}
\label{eqveff}
V_{eff}({\bf r},t)=V_n({\bf r})+\int \frac{\rho({\bf r}^{'},t)d^3r^{'}}{|{\bf r}-{\bf r}^{'}|}
+{\bf E}(t)\cdot {\bf r}+V_p({\bf r},t)
\end{equation}
where $V_n$ is the interaction with the nuclei, $\rho({\bf r},t)$ is the total electron density without the electron under consideration.
Note $V_p$ is absent in the SAO method.
The molecule is propagated from initial time, until the field reaches the maximum amplitude of 0.0755 a.u. Thus electronic dynamics are included in the effective potential.
The field-dressed and field-free orbital energy $\varepsilon$ of $5\sigma$ is presented in Fig.~\ref{veff} with horizontal marks.
$V_{eff}$ is asymmetric for $\beta=0^\circ$ [Fig.~\ref{veff}(a)] and $180^\circ$ [Fig.~\ref{veff}(b)], with a higher potential barrier for $\beta=180^\circ$.
If the field-free $\varepsilon$ is used, over-the-barrier ionization (OTBI) happens for $\beta=0^\circ$ and the original MO-ADK predicts a much smaller ionization ratio $P(180^\circ)/P(0^\circ)$ than unity.
The inclusion of polarization potential generally raises the potential barrier for both $\beta=0^\circ$ and $180^\circ$, leading to a suppression of the ionization probabilities in CS-ADK.
In our SAO method, $\varepsilon$ is lowered (lifted) for $\beta=0^\circ$ ($180^\circ$) due to the linear stark effects.
As a result, OTBI is almost satisfied for $\beta=180^\circ$ while electrons has to tunnel through a barrier to ionize for $\beta=0^\circ$, reversing the orientation-dependent ionization rate. The Stark corrected MO-ADK theory fails to explain experiments due to the same reason as the SAO method.
In the SAO+P method, although the potential barrier is raised for both orientations, $\varepsilon$ is shifted towards different directions: $\varepsilon$ is lifted for $\beta=0^\circ$ and lowered for $\beta=180^\circ$. It reflects the orbital distortion  and  also the dynamics of $5\sigma$ electrons by the dynamic core polarization.

\begin{figure}[t]
\centering
\includegraphics*[width=3.2in]{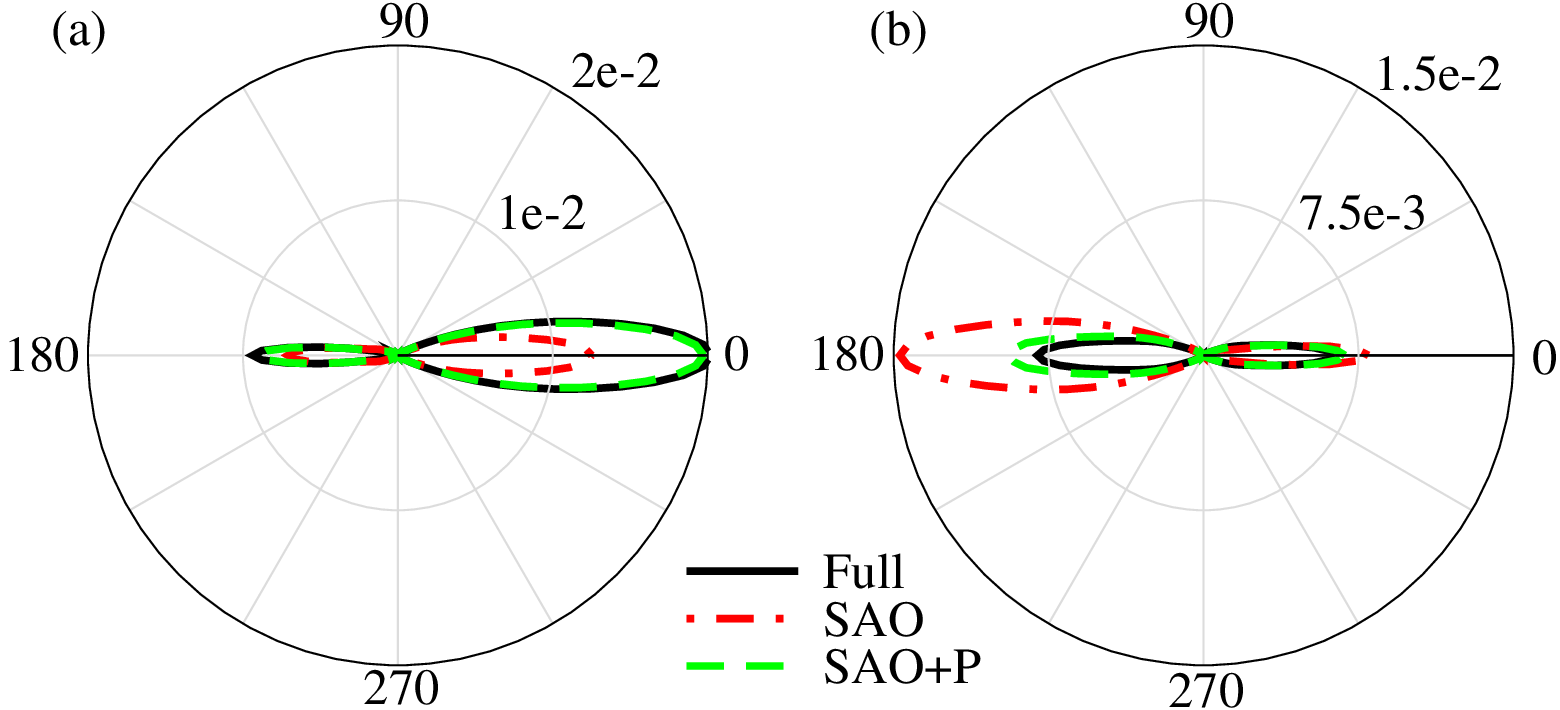}
\caption{(Color online)
Photoelectron angular distributions of CO for (a) parallel and (b) antiparallel orientation, computed by: the full method (solid lines), the SAO method (dot-dashed lines) and the SAO+P method (dashed lines) .
}
\label{pad3t}
\end{figure}

In order to fully characterize the ionization
dynamics,  we further investigate the photoelectron angular distribution (PAD) for different orientations.
The PAD in direction $\hat{\bf r}$ is calculated as~\cite{abu10}
\begin{equation}
\label{eqpad}
\frac{\partial P}{\partial \Omega}=\int_0^T \hat{\bf r}\cdot {\bf j}(R_b,t)dt
\end{equation}
where ${\bf j}(R_b,t)$ is the flux calculated at $R_b$ and time $t$ in direction $\hat{\bf r}=(\theta,\varphi)$.
The time $T$ is chosen long enough to allow all the flux to pass $R_b$.
In Fig.~\ref{pad3t}, the PADs integrated over the angle $\varphi$ are computed at the same laser parameters as the total ionization yields for $\beta=0^\circ$ and $180^\circ$.
It can be seen that the emission of photoelectrons are mainly directed along the molecular axis. The major loaf is ionized by the
peak electric field of the pulse, while subpeak fields ionize the minor part. The PADs from SAO method deviate from the full results just as the case of total ionizations (Fig.~\ref{iprob3t}). The SAO+P method yields
good agreement with the full method.
Together with Fig.~\ref{iprob3t} and Fig.~\ref{pad3t}, we identify that the dynamic core polarization does improve the original SAO method and is crucial for the correct description of ionization for multielectron molecules subjected to intense few cycle pulses.

In conclusion, we demonstrated that the orientation-dependence of strong-field ionization probabilities of CO  is essentially affected by the core electronic states of multielectron molecules.
The single active orbital method predicts qualitatively incorrect ionization yields due to the neglect of the core polarization dynamics.
By including the polarization potential from the laser polarized molecular core,  the results agree with the experiment allowing the identification of  the importance of the dynamic core polarization. It is expected to have  implications for high harmonic generation as well where the encoded multielectron effects are being actively explored \cite{mcfarlan08,smirnova09,patchkovskii06,zhao07pra}.
We conclude a theory beyond single-active-electron is in need for the tunneling ionization  of multielectron systems, by taking into account of the dynamical distortion of the ionizing orbital as a prerequisite.

This work is supported by  the National Basic Research Program of China (973 Program)
under grant no 2013CB922203, the NSF of China  (Grants No. 11274383) and the Major Research plan of NSF of China (Grant No. 91121017).
B. Z. is supported by  the Innovation Foundation of NUDT under Grant No. B110204
and the Hunan Provincial Innovation Foundation For Postgraduate
under Grant No. CX2011B010.



\end{document}